\documentclass[useAMS]{mn2e}
\usepackage{graphicx}
\title[EDOHS]{The rapid decay phase of the afterglow as the signature of the Blandford-Znajek mechanism}
\author[Nathanail et al.]{A. Nathanail$^{1,2}$\thanks{E-mail:
antonionitoni@hotmail.com}, A. Strantzalis$^{1,2}$
and I. Contopoulos$^{1}$ \\
$^{1}$Research Center for Astronomy and Applied Mathematics,
Academy of Athens, Athens 11527, Greece\\
$^{2}$Section of Astrophysics, Astronomy and Mechanics,
Department of Physics, University of Athens, \\
Panepistimiopolis Zografos, Athens 15783, Greece}

\def\gsim{\mathrel{\raise.5ex\hbox{$>$}\mkern-14mu
             \lower0.6ex\hbox{$\sim$}}}
\def\lsim{\mathrel{\raise.3ex\hbox{$<$}\mkern-14mu
             \lower0.6ex\hbox{$\sim$}}}

\begin{document}

\date{Accepted 1988 December 15. Received 1988
December 14; in original form 1988 October 11}

\pagerange{\pageref{firstpage}--\pageref{lastpage}} \pubyear{2015}

\maketitle

\label{firstpage}

\begin{abstract}
Gamma-ray bursts (GRBs) are believed to be powered by the
electromagnetic extraction of spin energy from a black hole
endowed with a magnetic field supported by electric currents in a
surrounding disk (Blandford \& Znajek~1977). A generic feature of
this mechanism is that, under certain fairly general assumptions,
the energy loss rate decays exponentially. In this work, we are
looking precisely for such exponential decay in the light curves
of long duration GRBs observed with the XRT instrument on the
Swift satellite. We found out that almost 30\% of XRT light curves
show such behavior before they reach the afterglow plateau.
According to Blandford \& Znajek, the duration of the burst
depends on the magnetic flux accumulated on the event horizon.
This allows us to estimate the surface magnetic field of a
possible progenitor. Our estimations are consistent with magnetic
fields observed in Wolf-Rayet stars.

\end{abstract}


\section[]{Introduction}

40 years after the discovery of Gamma-Ray Bursts (hereafter GRBs;
Klebesadel et al. 1973) their origin remains enigmatic. The
central engine, the photospheric emission, and the particular
emission mechanisms are still under debate (review Kumar \& Zhang
2014). It has been proposed that GRBs fall into two subcategories,
short- and long-duration (Kouveliotou et al. 1993), although
recently accumulated data suggests that this distinction may not
be as strong as originally thought (e.g. Ghirlanda, Nava \&
Ghisellini~2010). For long-duration GRBs the idea of the collapse
of a massive star and the formation of a stellar mass black hole
(Woosley 1993;
 Narayan, Piran \& Kumar 2001) is widely accepted.
A magnetar as central engine has also been proposed (Usov 1992),
and lately has been put into play for short GRBs too (Bucciantini
et al. 2011). Currently, the most popular model for short GRBs is
the merging of two compact objects (Rezzolla et al. 2011).

If the central engine is a black hole, there are two main physical
mechanisms that may power the burst: neutrino annihilation (e. g.
Chen \& Beloborodov 2007), and/or the extraction of the rotational
energy from a black hole (Lee et al. 2000). The first mechanism
may work in some cases but fails to explain the observed
energetics of Ultra Long GRB (Leng \& Giannios 2014). The second
involves the electromagnetic extraction of energy from the black
hole rotation (the Blandford-Znajek mechanism, hereafter BZ;
Blandford \& Znajek~1977), and is widely discussed in the GRB
literature (Lee et al. 2000, Wang et al. 2002, McKinney 2005,
Komissarov \& Barkov~2009, Nagataki~2009).

Any discussion of potential GRB models lead us to search for their
signatures in GRB observations, and in particular in X-rays where
a treasure load of data from the BAT and XRT instruments aboard
the Swift satellite are available. X-ray GRB light curves consist
of two different components: the prompt and the afterglow emission
(see Fig.~1). The prompt emission is triggered by BAT, and ends
when it decays below the instrument's sensitivity limit.
Anything below that limit is considered as afterglow emission
which itself consists of two components separated
phenomenologically (O~Brien et al.~2006; Willingale et al.~2007;
Ghisellini et al.~2009). The first afterglow component seems to
consist the extension of the prompt emission rapid decay phase,
noticed in the first X-ray afterglow detections of Swift
(Tagliaferri et al. 2005). This early steep decay phase was
discussed already before Swift (Kumar \& Panaitescu 2000). It was
originally modelled as a residual off-axis (or high latitude)
emission, and several others studied whether that model fits the
observations. According to this scenario, the emission follows a
`steepening' power law. As we will show below, in many cases, this
early steep decay phase may be fit by a single exponential. The
second afterglow component starts with a plateau, continues in a
power law phase, and ends with a final fast decay. The theoretical
understanding for this second component involves energy loss
through external shocks in the expanding post-explosion stellar
material (Meszaros \& Rees 1997).

If we believe that BZ is the mechanism that powers the burst, we
should be able to find its signature in the X-ray GRB light curves
which allow us to follow the central engine activity for very long
times (Zhang et al. 2014). Obviously, as the black hole loses
energy, it will spin down. The timescale of this procedure depends
on the magnetic flux accumulated on the event horizon of the black
hole and the angular velocity of the event horizon (Blandford \&
Znajek~1977; Lee et al. 2000). If the accumulated magnetic field
exceeds magnetar values ($10^{15}$~G), the spindown of a stellar
mass black hole could last for tens of seconds. Unfortunately, due
to the time it takes to reposition the satellite after the BAT
triggering in order for the (much more sensitive) XRT instrument
to begin observing the burst, the first 10 to 100 seconds of this
phase are most of the time not observed. We will here show that,
in about 30\% of XRT observations where the first rapid decay
afterglow phase is clearly observed, parts of it follow closely
the theoretical curve (exponential) dictated by the BZ black hole
spindown!

 \begin{figure}
\includegraphics[width=75mm]{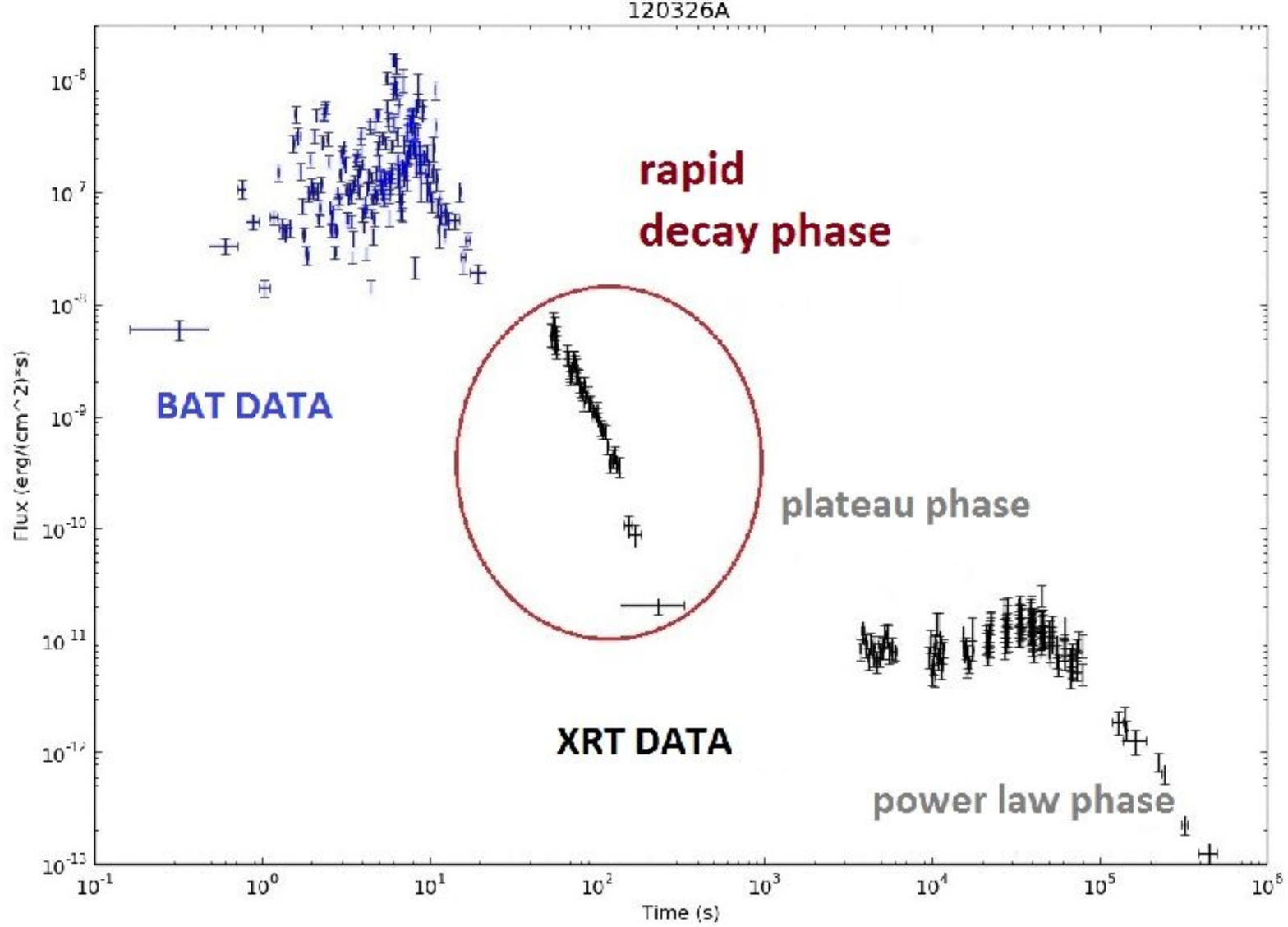}
\caption{Lightcurve of $GRB ~120326 A$.  Log-Log plot. Up left (in
blue) are the BAT data, whereas the rest (in black) is the X-ray
afterglow from XRT. We are mostly interested in the first part of
the afterglow, the so-called rapid decay phase, which is suspected
to be of internal origin (central engine activity). In a large
number of bursts, this rapid decay phase seems to follow an
exponential, compatible with black hole spindown.}
 \label{Cartoon}
 \end{figure}

In Nathanail \& Contopoulos~(2015) we showed that in the limit of
slow black hole rotation (i.e. less than about 10\% of maximum
rotation) a simple exponential can describe the energy loss rate
of the BZ mechanism. We thus associated this with Ultra Long GRBs,
and argued that what sets the duration of the central engine
activity is the amount o magnetic flux accumulated on the event
horizon of the newly formed black hole after the core collapse of
a suppermassive star. Thus, the surface magnetic field of the
progenitor star could be crucial for the central engine activity.
Based on these ideas we suggested that Ultra Long GRBs lie in the
same class together with the usual Long GRBs, and their
extraordinary duration may just be due to the low surface magnetic
field of the progenitor star.

The present study is a follow up of the previous one, in the sense
that it allows us to extend these ideas to the full GRB
population. If our magnetic field estimations have anything to do
with reality, this has a straightforward implication that GRB
outflows are magnetized (or at least the ones that are described
by the BZ mechanism). In \S~2 we give the basic equations and
discuss the physical problem. In \S~3 we show our results for a
large sample of Swift GRBs where we focus on XRT observations that
adequately follow the central engine activity. In \S~4 we discuss
how we can use this information to estimate the magnetic field in
the vicinity of the event horizon of the black hole, and from
there link it with the surface magnetic field of a possible
progenitor star. Finally, we end in \S~5 with a summary of our
work.

\section[]{Black Hole Spin Down}

Let us consider a supermassive progenitor star whose core
collapses and forms a rotating black hole. It is natural for the
star to be magnetized. Highly conducting matter from the interior
of the star will drive the advection of magnetic flux during the
collapse. A certain amount of magnetic flux $\Psi_m$ is then going
to cross the horizon. An equatorial  thick disk (torus) will form
around the black hole due to the rotational collapse. A black hole
cannot hold its own magnetic field, but the material from the
thick disk will act as a barrier that will hold the magnetic flux
initially advected.

As long as this is the case, the black hole will lose
rotational/reducible energy at a rate
\begin{equation}
\dot{E} \approx -\frac{1}{6\pi^2 c}\Psi_m^2\Omega^2\ ,
\label{EdotIa}
\end{equation}
and will thus spin down very dramatically (Blandford \& Znajek
1977 for low spin parameters; Tchekhovskoy et al. 2010,
Contopoulos et al. 2013, Nathanail \& Contopoulos 2014 for
maximally rotating black holes). $\Omega$ is the angular velocity
of the black hole horizon. In principle, this procedure can
extract almost all the available/reducible energy $E_{\rm rot}$
(Christodoulou \& Ruffini 1971; Misner, Thorne \& Wheeler~1973).
The reader can check the above references to see that the
rotational energy of a $10M_{\odot}$ initially maximally rotating
black hole is equal to
\begin{equation}
E_{\rm max\ rot}=29\% Mc^2 \approx 5\times 10^{54}\ \mbox{erg}\ ,
\end{equation}
a rather extreme value for the total energy released in a GRB
explosion (Komissarov, personal communication). However, if the
black hole is e.g. rotating at $10\%$ of maximum, then
\begin{equation}
E_{\rm 10\%\ of\ max\ rot}\approx 2\times 10^{52}\ \mbox{erg}\ ,
\end{equation}
which is much more reasonable for a GRB. It is clear that if we
change the mass and the spin of the black hole, the energy it can
give off spans more than three orders of magnitude.

 \begin{figure}
\includegraphics[width=84mm]{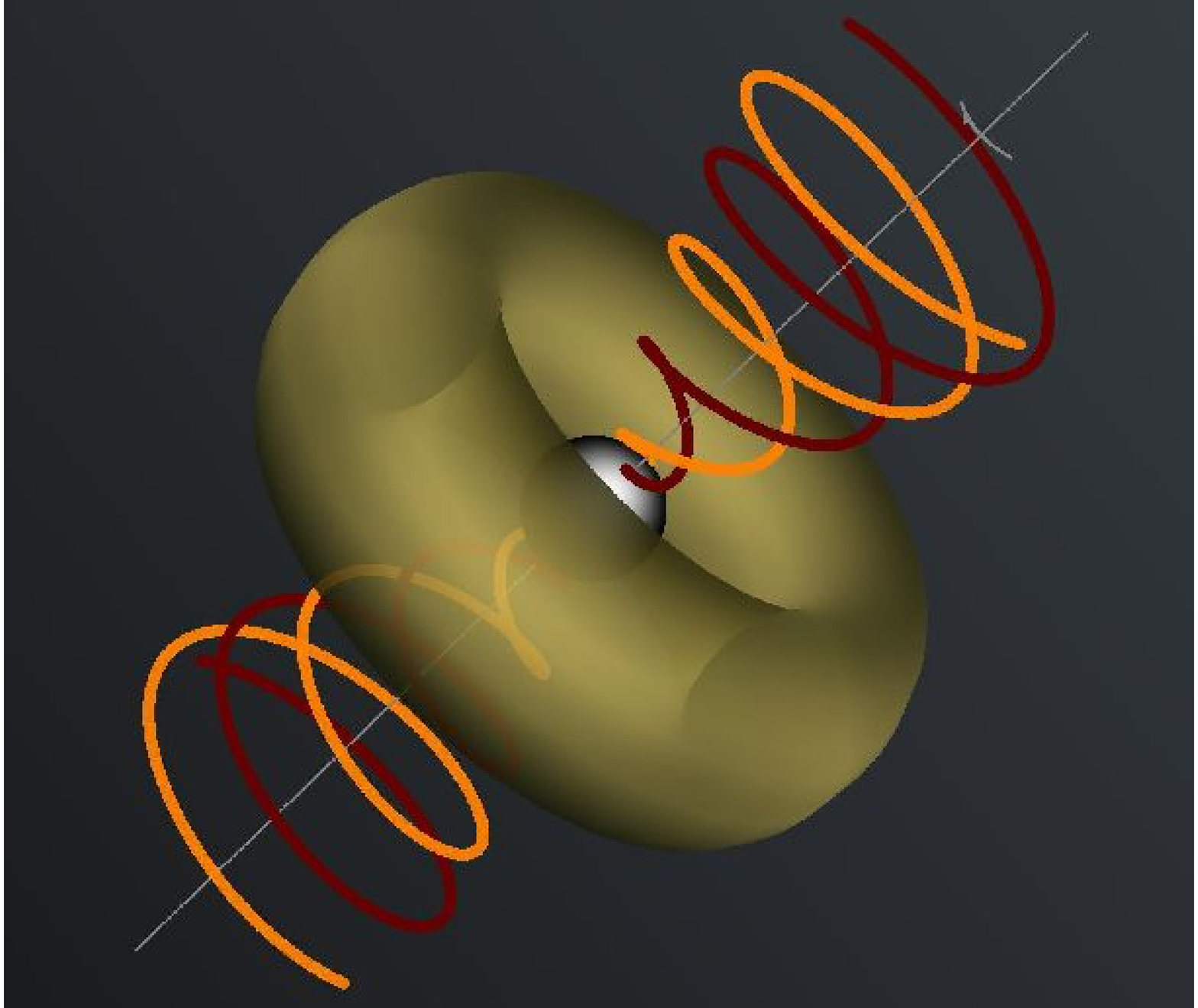}
\caption{The structure of the magnetosphere close to the event
horizon of a rotating black hole. Magnetic field lines (depicted
in  dark red and orange) based on the solutions of Nathanail \&
Contopoulos~(2014). A massive torus of material (transparent)
holds the magnetic flux on to the event horizon.}
 \label{BHTRYLAST}
 \end{figure}

In what follows, we will assume that the newly formed
black hole is slowly rotating. Under that approximation,
$M\approx$~const. and
\begin{equation}
E_{\rm rot} \approx \frac{1}{8}Mc^2\left(\frac{\Omega}{\Omega_{\rm
max}}\right)^2\ , \label{Erot}
\end{equation}
where $\Omega_{\rm max}\equiv c^3/2{\cal G} M$ is the angular
velocity of a maximally rotating black hole, and ${\cal G}$ is the
gravitational constant. The black hole will, therefore, spin down
as
\begin{equation}
\dot{E} = \frac{{\cal G}^2M^3}{2c^4}\frac{{\rm d}(\Omega^2)}{{\rm
d}t}\ .\label{EdotIb}
\end{equation}
Equating eqs.~(\ref{EdotIa}) and (\ref{EdotIb}) and solving for
$\Omega^2=\Omega^2(t)$, we obtain
\begin{equation}
\dot{E} \propto e^{-t/\tau_{\rm BZ}} \label{Eapprox}
\end{equation}
where
\begin{equation}
\tau_{\rm BZ}\equiv \frac{3 c^5}{16{\cal G}^2 B^2 M}=50
\left(\frac{B}{10^{15}\ \mbox{G}}\right)^{-2} \left(\frac{M}{10
M_\odot}\right)^{-1}\mbox{sec}\ \label{tBZ}
\end{equation}
is the timescale for the spinning down procedure. We have defined
here a typical value for the accumulated black hole magnetic field
required by the Blandford-Znajek mechanism,
\begin{equation}
B =\frac{\Psi_m c^4}{4\pi {\cal G}^2 M^2}\ .
\end{equation}
$B$ can reach very high values during the core-collapse of a
massive star, and for $B\sim 10^{15}$~G, the black hole spins down
in less than a hundred seconds. It is interesting to notice here
that eq.~(\ref{Eapprox}) cannot distinguish between a black hole
and a neutron star/magnetar with field lines that are held open by
the surrounding material. Therefore, in principle, we cannot claim
that exponential decay GRBs is definite proof for the presence of
a central black hole (it is the same as pulsar spindown with
breaking index equal to unity).  It is, however, a strong
suggestion since, if a cavity forms around the central object,
magnetic field lines will `prefer' to form a dead zone of closed
lines instead of extending to infinity. This effect will lead to a
pulsar spindown different from the exponential one that we are
investigating in this work. On the other hand, the exponential
decay suggest the presence of very high magnetic fields that can
drive an electrodynamic spindown.

It is usually argued that the balance of magnetic pressure with
ram pressure from the disk can give an estimate of the possible
magnetic flux accumulated around the central object. This argument
is reasonable for accreting black hole systems such as AGNs and
X-ray binaries, {\em but not} in GRB events where the black hole
forms inside a supper massive star. In the latter, it is very
reasonable for this magnetic field to be held in place by a
massive disk/torus of material that {\em does not accrete}. A
crude calculation of the force balance between the outward
electromagnetic force, gravity and rotation yields
\begin{equation}
\frac{B^2}{r}r^3 \sim \frac{G M M_{\rm d}}{r^2} - \frac{M_{\rm d}
l_{\rm d}^2}{r^3} \label{Bdisk}
\end{equation}
where, M$_{\rm d}$ is the mass, $l_{\rm d}$ is angular momentum
per unit mass, and $r$ is the radius and approximate height of the
torus. If the disk is rotationally supported, eq.~(\ref{Bdisk})
does not allow for any extra magnetic field to be held in its
interior. This could be the case for a progenitor star with
relatively fast rotation. If, on the other hand, the progenitor
star is not rotating as fast, a slowly rotating black hole may
form at the center (as we argued above), while the rest of the
left over stellar material may not have enough angular momentum to
form a centrifugally supported disk around it (Woosley \& Heger
2006, 2012). In that case, it is natural to imagine that the
equilibrium described by eq.~(\ref{Bdisk}) is reached. One can
easily check that, in order to support a magnetic field strength
of $B\sim 10^{15}$~G for very small values of $l_{\rm d}$, a torus
of size $r\sim 2{\cal G}M/c^2$ and mass $M_d\sim
10^{-5}~M_{\odot}$ around a $10M_{\odot}$ black hole is all that
is needed. For higher values of $l_{\rm d}$ one needs a higher
torus mass to hold the same value of the magnetic field. Notice
that we are not presently considering the stability of this
configuration against e.g. Rayleigh-Taylor instability
(Contopoulos \& Papadopoulos 2012). We just assume that it
survives for the duration of the black hole spindown that we
propose we are observing in a GRB. Obviously, if the massive disk
is dispersed faster than the duration of the spindown, the
accumulated magnetic flux $\Psi_m$ will not be conserved, and the
spindown evolution will not be exponential.

Other effects may too modify the black hole electromagnetic
spindown, making it difficult to discern its activation and
evolution. GRB events  may be `contaminated' by extra events that
possibly take place during the spindown. One example may be fall
back accretion of huge amounts of mass that could lead to a spin
up of the black hole with a subsequent different spindown
evolution. Moreover, the electromagnetic interaction with the
torus formed around the black hole may result in an extra spindown
that may too be linked to GRBs (van Putten et al. 2009).


\section[]{XRT data}

Up to this point we have shown that rotating black holes embedded
in a strong fixed magnetic field spin down almost exponentially.
It is, therefore, natural to search for exponential decay in the
light curves of GRBs.

 \begin{figure*}
\includegraphics[width=184mm]{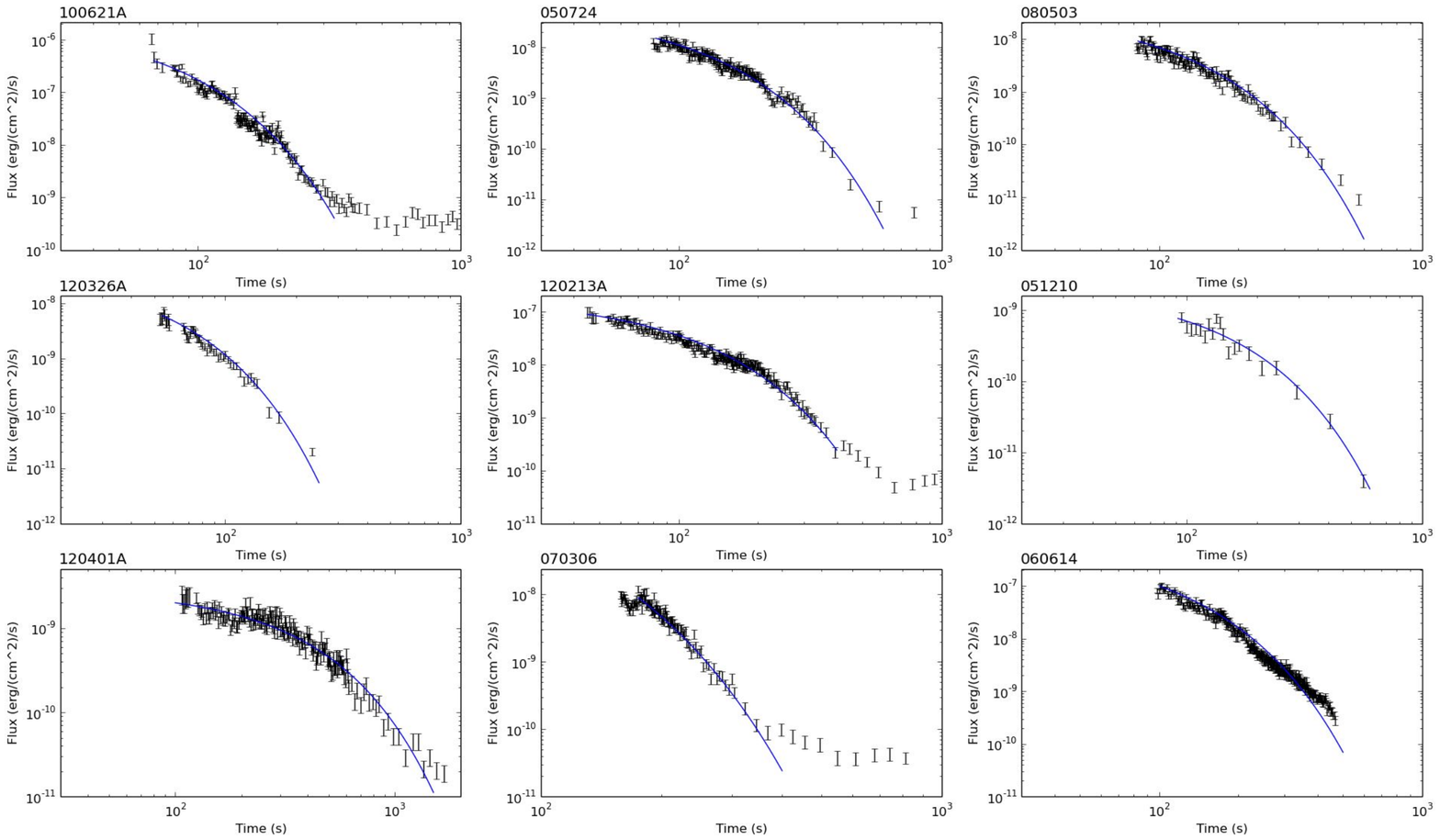}
\caption{Nine characteristic GRBs.  Log-Log plot. The green curve
is the theoretical exponential black hole spindown. Energy flux
at $0.3~-~10$~keV. We focus in the first part of the afterglow,
the rapid decay phase.}
 \label{ULGRB3}
 \end{figure*}

Discerning the central engine activity in a GRB light curve  can
be tricky. The long-term activity of the central engine must be
identified. In that respect, the time estimate $T_{90}$ can be
misleading. When the signal drops out of the $\gamma$-ray band, it
continues in the X-rays and this emission can still be of an
internal origin. A different duration of the burst must,
therefore, be defined. This was done recently by Zhang et
al.~(2014) who completed a comprehensive study of Swift XRT light
curves that show the extended central engine activity. For all the
bursts in their sample a $t_{\rm burst}$ is assigned. Till that
time, the emission can be argued to be of internal origin
(dominated by emission from a relativistic jet via an internal
dissipation process), not dominated by the afterglow emission from
the external shock.

We used the same GRB sample with Zhang et al. 2014. This is
because we want to focus in the X-rays and we need XRT light
curves with enough data to follow the central engine activity. All
the XRT light curves are taken from the Swift/XRT
website\footnote{http://ww.wift.ac.uk/xrt.curves/ } (Evans et al.
2009)  at the UK Swift Science Data Centre (UKSSDC). Our aim is to
check for signs of exponential decay. While the $\gamma$-ray
signal drops we are left  with X-rays. Thus, in most GRBs we can
follow the long term evolution of the burst in X-rays.  At first
the energy flux shows (in most cases) a very steep decay which is
believed to be the tail of the prompt $\gamma$-ray emission. Then
it enters a plateau phase and continues as a power law (Wijers et
al.~1997). This extended emission can some times last up to a few
weeks and is most probably associated with an external origin
(external shocks, e.g. Meszaros \& Rees 1997). Our aim is to check
for signs of exponential decay in the first steep decay phase. It
has been suggested that this rapidly declining X-ray light curve
shows the evolution of the central engine activity with time (Fan
\& Wei 2005). We agree with this interpretation  and we believe
that when we find exponential decay in this first (steep) phase we
are observing the evolution of the black hole. Moreover, we argue
that this may be associated with the black hole spindown discussed
in \S~2.

We considered every light curve from our sample and tried to fit
eq.~(\ref{Eapprox}) over some part of it. This fit allowed us to
estimate $\tau_{\rm BZ}$, which is a really important physical
parameter. Knowing $\tau_{\rm BZ}$, we can estimate the strength
of the magnetic field in the vicinity of the black hole. We check
the quality of our fit with an R-squared statistics parameter (the
closer R is to 100\%, the better our exponential fits the data).

As discussed above, we want to find  exponential decay in the
first steep decay phase of  the light curves. In order to
guarantee good statistical results, we want to be able to follow
this decay for more than one order of magnitude, so that we are
confident we are following an exponential and not just a steep
power law. Cases where  data points are too few to claim good statistical results
(e.g. GRBs 090418A, 130803A, 051016A, 090313 etc.) were excluded.
For the ones that the rapid decay in the X-rays is less than one
order of magnitude (e.g. GRBs 100219A, 091020, 090904B, 090515,
etc.) we may have lost the first afterglow phase and thus are left out.
Furthermore, GRBs with an irregular distribution of data points
(e.g. GRB 050915A) (i.e. with extra complex physical process going
on in parallel) are left undefined.

 \begin{figure*}
\includegraphics[width=184mm]{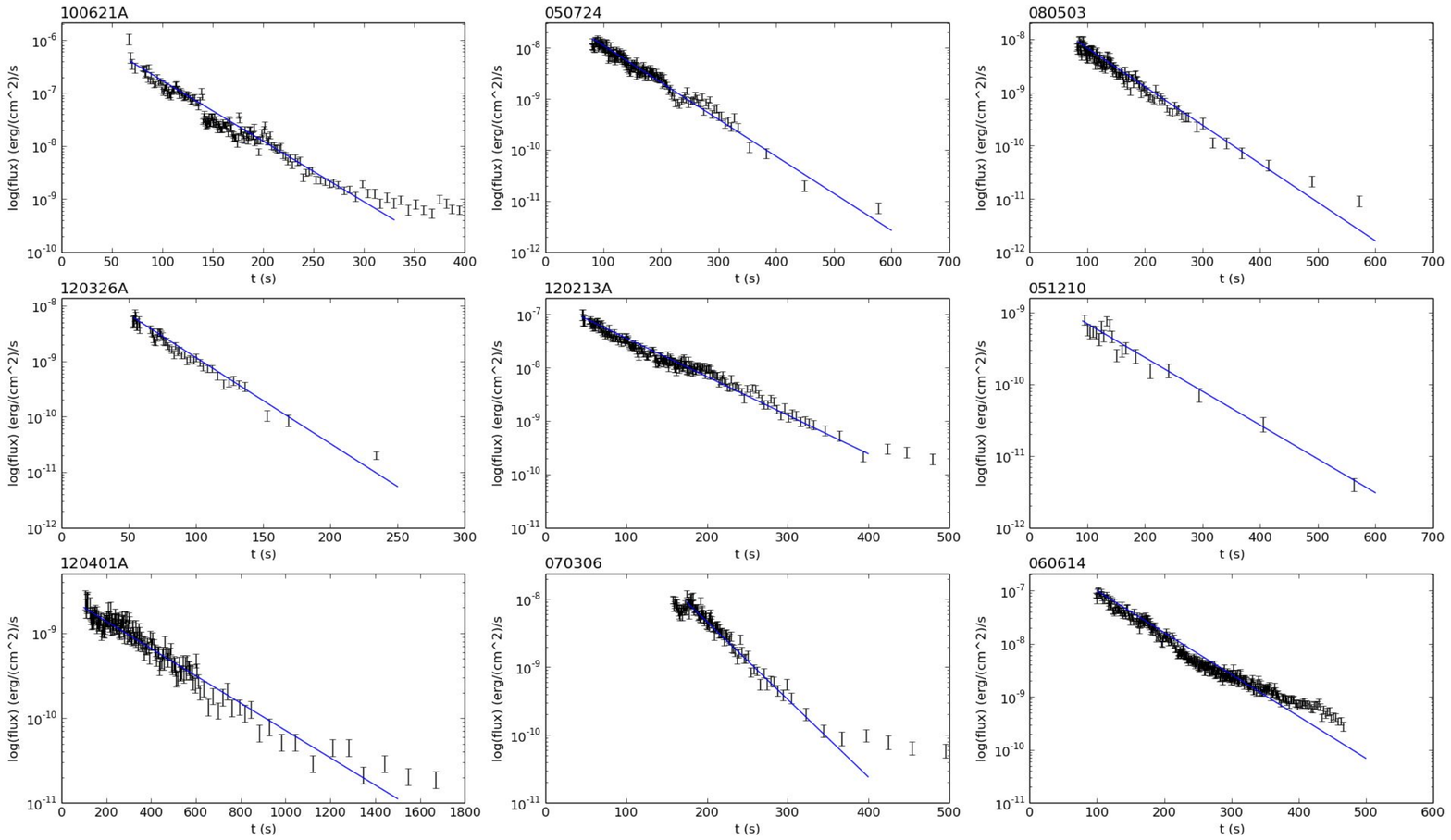}
\caption{Same as Fig.~3 in Log-Linear scale. Notice that in Log-Linear plots  an
   exponential is shown as a straight line.
Energy flux  at $0.3~-~10$~keV. We focus in the first part of the afterglow, the rapid decay phase. }
 \label{ULGRB3}
 \end{figure*}

All these features discussed here have been categorized in a
phenomenological manner which identified three components in the
afterglow: rapid decay followed by a plateau and a final power law
phase (Ghisellini et al. 2009). In this way, we can identify
several X-ray flares that require extended central engine
activity. Going back to our theoretical model we remind the reader
that many effects (such as mass infalls that may result in sudden
black hole spin ups) can modify the black hole electromagnetic
spindown, making it difficult to discern its activation and
evolution. Such secondary events will yield secondary flares, as
is frequently seen in GRB ligh curves(Wu et al. 2013). There are
cases in which, after a big flare, the  emission shows a steep
decay more than two orders of magnitude (e.g.  GRB 121027A). It is
very interesting that this rapid emission decay after the flare is
also exponential. We have found several events that conform to
this picture, and we plan to discuss them in more detail
elsewhere. There are also some GRBs with minor flares that do not
disrupt a single exponential fit. In these GRBs R-squared is less
than $70-80\%$ (e.g. GRB 111103B) and its not clear whether we are
following a black hole spinning down or not. Another physical
possibility that would mask the exponential decay is if the
surrounding disk disperses faster than the duration of the
spindown. In that case, the accumulated magnetic flux $\Psi_m$
will not be conserved, and the spindown evolution will not be
exponential.
 \begin{figure}
\includegraphics[width=89mm]{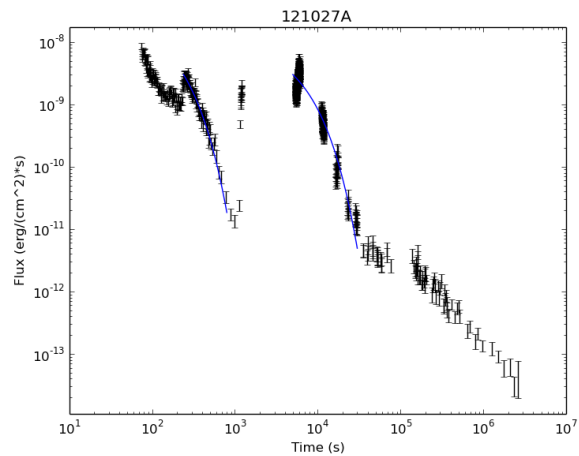}
\caption{GRB 121027A, a huge X-ray flare made this one
exceptional. It is interesting to see that after the flare, the
decay is exponetial.}
 \label{Flare}
 \end{figure}
We decided to focus only on GRBs with a single emission decay
event. We thus formulated the following empirical criteria that
characterize our good sample: (i) emission decay more than one
order of magnitude, (ii) full (not sparse) sampling of the light
curve in the time interval  that we follow the spindown and (iii)
the steep decay emission phase to be  without continuous big
flares. No strong conclusions on the central engine activity can
be extracted for those few GRBs ($\sim 33$) with not enough data
in the steep decay phase.

By fitting eq.~(\ref{Eapprox}) to the light curve, we find
$\tau_{\rm BZ}$ for every GRB listed in tables~1 \& 2. From the
primary sample of 343 GRBs, we excluded $33$ objects with low
number of data in the rapid decay phase, and were left with $310$
GRBs. From these GRBs, $60$ ($\sim 20\%$) have a very good
exponentially decaying emission event, with$~R$- squared over
$90\%$ for most of them. This can be our golden sample. In  table
1 we show the $\tau_{\rm BZ}$ and the R-squared that we obtained
for these objects. In all these events we can follow the
exponential decay (black hole spindown) for more than two orders
of magnitude combined with a very good fitting result (R-squared
more than $90\%$). There are many GRBs ($\sim 31$) where the rapid
decay phase is really small and the lightcurve enters quickly in a
plateau phase or go straight to the plateau phase ($\sim 33$). In these
events, the rapid decay can be less than an order of magnitude,
thus we cannot support an exponential.

\begin{table*}
 \centering
 \caption{The 60 GRBs with a clear exponential decay}

\begin{tabular}{ccc|ccc|ccc}

 \hline

   \multicolumn{9}{|c|}{Fitted Parameter} \\

\hline

{\em GRB} & {\em $\tau_{\rm BZ}$} (sec) &{\em R-squared} &{\em GRB} & {\em $\tau_{\rm BZ}$} (sec) &{\em R-squared}
 &{\em GRB} & {\em $\tau_{\rm BZ}$} (sec) &{\em R-squared}\\

 \hline
$050716$ & $140~(\pm 4)~$ & $0.929$  & $050724$ & $60~(\pm 4)~$ & $0.948$ & $050915B$ & $31~(\pm 4)~$ & $0.916$\\ $051210$ & $90~(\pm 8)~$ & $0.741$  & $060413$ & $82~(\pm 13)~$ & $0.800$ & $060614$ & $55~(\pm 2)~$ & $0.988$\\ $060708$ & $25~(\pm 3)~$ & $0.945$  & $060729$ & $35~(\pm 3)~$ & $0.975$ & $061110A$ & $50~(\pm 4)~$ & $0.921$\\ $061121$ & $14~(\pm 1)~$ & $0.973$  & $061222A$ & $34~(\pm 4)~$ & $0.849$ & $070306$ & $38~(\pm 3)~$ & $0.928$\\ $070420$ & $36~(\pm 3)~$ & $0.896$  & $070621$ & $50~(\pm 6)~$ & $0.897$ & $071227$ & $72~(\pm 20)~$ & $0.726$\\ $080205$ & $32~(\pm 4)~$ & $0.959$  & $080229A$ & $33~(\pm 5)~$ & $0.940$ & $080503$ & $60~(\pm 4)~$ & $0.920$\\ $081028$ & $145~(\pm 25)~$ & $0.830$  & $081128$ & $50~(\pm 5)~$ & $0.973$ & $081221$ & $32~(\pm 2)~$ & $0.918$\\ $081230$ & $17~(\pm 2)~$ & $0.958$  & $090111$ & $28~(\pm 2)~$ & $0.929$ & $090404$ & $28~(\pm 2)~$ & $0.972$\\ $090618$ & $18~(\pm 2)~$ & $0.904$  & $091026$ & $40~(\pm 3)~$ & $0.925$ & $091029$ & $30~(\pm 2)~$ & $0.936$\\ $091104$ & $70~(\pm 10)~$ & $0.801$  & $100418A$ & $30~(\pm 3)~$ & $0.948$ & $100425A$ & $25~(\pm 3)~$ & $0.826$\\ $100514A$ & $32~(\pm 2)~$ & $0.944$  & $100522A$ & $18~(\pm 3)~$ & $0.965$ & $100526A$ & $57~(\pm 5)~$ & $0.949$\\ $100615A$ & $26~(\pm 3)~$ & $0.945$  & $100621A$ & $38~(\pm 2)~$ & $0.989$ & $100725A$ & $95~(\pm 7)~$ & $0.871$\\ $101030A$ & $36~(\pm 3)~$ & $0.926$  & $101213A$ & $75~(\pm 9)~$ & $0.902$ & $101225A$ & $6000~(\pm 0)~$ & $0.900$\\ $110210A$ & $90~(\pm 8)~$ & $0.927$  & $110414A$ & $30~(\pm 6)~$ & $0.906$ & $110420A$ & $26~(\pm 2)~$ & $0.939$\\ $110808A$ & $50~(\pm 6)~$ & $0.946$  & $111123A$ & $130~(\pm 6)~$ & $0.968$ & $111209A$ & $4900~(\pm 500)~$ & $0.900$\\ $111225A$ & $110~(\pm 18)~$ & $0.826$  & $120106A$ & $21~(\pm 1)~$ & $0.929$ & $120116A$ & $39~(\pm 6)~$ & $0.900$\\ $120213A$ & $60~(\pm 5)~$ & $0.982$  & $120215A$ & $65~(\pm 8)~$ & $0.982$ & $120324A$ & $44~(\pm 4)~$ & $0.902$\\ $120326A$ & $28~(\pm 2)~$ & $0.959$  & $120401A$ & $270~(\pm 30)~$ & $0.917$ & $120514A$ & $30~(\pm 4)~$ & $0.909$\\ $120922A$ & $75~(\pm 3)~$ & $0.960$  & $121108A$ & $15~(\pm 2)~$ & $0.936$ & $130315A$ & $80~(\pm 5)~$ & $0.983$\\ $130528A$ & $25~(\pm 2)~$ & $0.924$  & $131018A$ & $50~(\pm 13)~$ & $0.794$ & $131127A$ & $30~(\pm 5)~$ & $0.940$
\end{tabular}
\end{table*}

In table 2 we list those $23$ objects that have flares with a
subsequent exponential decay. A representative example from this
class is GRB 121027 (shown in fig.~5) were the X-ray lightcurve
starts with an exponential (not disrupted by a small flare around
$\sim 300$~sec), and after a big flare at around $\sim 5000$~sec
(energy flux increases more than two orders of magnitude) it
continues again with an exponential. This big flaring activity may
be understood as black hole spinning up because of large amount of
mass infall. Altogether,  $27\%$(83) of GRBs show an exponential
decay in the first phase of the afterglow. There are another $35$
GRBs with many small flares in their lightcurves, for which no
conclusion is reached.

We have to state here that we found  $28$ GRBs in which the whole
afterglow can be fitted with a single power law.  For the remaining $100$ 
objects no strong conclusion could be reached.
 There is a possibility that XRT failed to
catch their rapid decay phase. Moreover, a further spindown of the
central object may be hidden behind a stronger emission of
external origin.

\section[]{The magnetic field of the progenitor star}

In the previous  section we showed that there are several GRBs
with clear signs of exponential decay in their light curves. We
have related this decay to the spindown of a newly formed black
hole (most probably slowly rotating). The spindown is
electromagnetic, and  the timescale $\tau_{\rm BZ}$ gives us an
estimate of the magnetic field strength on the event horizon of
the black hole.

Some GRBs were reported with ultra long central engine activity
(Levan et al. 2014). In this small population we also found clear
signs of exponential decay which we associated with black hole
spindown (Nathanail \& Contopoulos 2015). We were thus able to
link the magnetic flux accumulated on the event horizon to the
surface magnetic field of the progenitor star. It is natural to
extend this discussion to the long duration GRBs of our present
sample. As before, we will assume that a $10 M_{\odot}$ black hole
forms at the center. This is a natural choice if the progenitor
star mass is $25~-~40~M_{\odot}$ (Heger et al. 2003).
The timescales $\tau_{\rm BZ}$ that we have found span a range
between 11 and 6000~sec. Notice that these values are not
corrected for cosmological redshift. Applying them to
eq.~(\ref{tBZ}), we find that the magnetic field (uncorrected for
redshit) on the event horizon varies between
\begin{equation}
B_{H} \approx 10^{14}\ \ \mbox{and}\ \ 10^{15}~{\rm G}.
\end{equation}
This magnetic field will drive the black hole spindown, in
agreement with observations that show signs of magnetically
dominated outflows from GRBs (e.g. Guiriec et al. 2014).
\begin{table*}
 \centering
 \caption{The 23 GRBs with a clear exponential decay after a flare}

\begin{tabular}{ccc|ccc|ccc}

 \hline

   \multicolumn{9}{|c|}{Fitted Parameter} \\

\hline

{\em GRB} & {\em $\tau_{\rm BZ}$} (sec) &{\em R-squared} &{\em GRB} & {\em $\tau_{\rm BZ}$} (sec) &{\em R-squared}
 &{\em GRB} & {\em $\tau_{\rm BZ}$} (sec) &{\em R-squared}\\

 \hline
$050502B$ & $95~(\pm 8)~$ & $0.928$ & $060929$ & $90~(\pm 14)~$ & $0.899$ &  $061202$ & $55~(\pm 3)~$ & $0.923$  \\
$070720B$ & $55~(\pm 7)~$ & $0.905$ & $080212$ & $47~(\pm 4)~$ & $0.944$ &  $080325$ & $95~(\pm 8)~$ & $0.927$   \\
$090621A$ & $27~(\pm 3)~$ & $0.921$ & $090904$ & $40~(\pm 6)~$ & $0.952$ &  $100619A$ & $11~(\pm 1)~$ & $0.965$   \\
$100704A$ & $36~(\pm 3)~$ & $0.960$ & $100727A$ & $45~(\pm 6)~$ & $0.911$ &  $100802A$ & $110~(\pm 17)~$ & $0.910$  \\
$100814A$ & $65~(\pm 5)~$ & $0.936$ & $100902A$ & $37~(\pm 3)~$ & $0.981$ &  $100906A$ & $12~(\pm 1)~$ & $0.985$  \\
$120308A$ & $52~(\pm 4)~$ & $0.910$ & $121027A (1)$ & $110~(\pm 6)~$ & $0.929$ &  $121027A (2)$ & $3800(\pm 300)~$ & $0.700$  \\
$121123A$ & $110~(\pm 12)~$ & $0.979$ & $121217A$ & $80~(\pm 8)~$ & $0.917$ &  $131030A$ & $25~(\pm 2)~$ & $0.948$  \\
$140108A$ & $12~(\pm 2)~$ & $0.944$ & $140114A$ & $65~(\pm 4)~$ & $0.941$
\end{tabular}
\end{table*}
The black hole event horizon for a slowly rotating $10 M_{\odot}$
black hole is at $r_{H} \approx 3\times 10^6$~cm. During the
collapse, the magnetic field is carried along by the conducting
matter of the stellar interior, and flux conservation implies
\begin{equation} B r^2 = B_{\star} r_{\star}^2\ ,
\label{Bstar1}
\end{equation}
where $B_{\star}$ and $r_{\star}$ are the surface magnetic field
and the radius of the star respectively. A typical radius for  a
Wolf-Rayet star is $10^{12}$~cm (Crowther 2007), in which case
eq.~(\ref{Bstar1}) yields
\begin{equation}
B_{\star}\sim 10^3~-~10^4~{\rm G}\ . \label{Bstar}
\end{equation}

Let us now see how the above estimates compare with observations
of magnetic fields in Wolf-Rayet stars. At visible wavelengths,
their stellar surface is hidden by a dense nebula. Magnetic field
values of $22 ~ - ~ 128$~G have been reported in their stellar
winds through measurements of emission lines (de la Chevrotiere et
al. 2014). The corresponding surface value of the magnetic field
must be much higher than the observed estimated values. Assuming
that the surface field is stretched as $1/r^2$ with distance in
the wind, our magnetic field estimate of eq.~(\ref{Bstar}) ten
stellar radii from the surface would yield 10 to 100~G, in
agreement with the observations.
Notice that our magnetic field estimates did not take into account
a $(1+z)^{1/2}$ correction factor due to the cosmological redshift
$z$. Notice that they also did not take into account possible
dynamo magnetic field amplification under the cataclysmic
conditions in the collapsing environment, as discussed in the
literature (Obergaulinger et al. 2009). If we assume an extra
three orders dynamo field amplification our estimate of the
surface magnetic field could be as low as $B_{\star}\sim 1$~G.

The above estimates were obtained with the physical image of a
Wolf-Rayet star discussed extensively in the GRB literature
(Woosley \& Bloom 2006). Even if the type of the progenitor
changes, our estimate of its magnetic field still holds and can be
slightly corrected (depending on the radius of the star). The main
point here is that the duration of these bursts depends on the
magnetic field  of the progenitor star. The idea that magnetic
flux is  the principal parameter that sets the luminosity of a GRB
is discussed also in Tchekhovskoy \& Giannios (2015), although in
their case the central engine  turns-off when the steep decline
stage starts.

\section[]{Summary}

We have shown that a newly formed black hole resulting from the
core collapse of a supermassive star may be slowly rotating and
still power a GRB. The magnetic flux accumulated on the event
horizon of the black hole can drive the extraction of the black
hole rotational energy through the Blandford-Znajek mechanism. We
argue that it is possible to hold this magnetic flux for the
duration of the burst even without accretion onto the black hole.
In that case, the electrodynamic energy release decays
exponentially. 

In the present study we extended the analysis of Nathanail \&
Contopoulos~(2015) on Ultra Long GRBs to a much larger sample of
Long GRBs. As before, in order to follow the central engine
activity as far as possible, we focused on the XRT X-ray band.
From a sample of $292$ GRBs, $60$ show a clear exponential decay
and $25$ also proceed with an exponential decay after a flare. In
total, $29\%$ of the events in our sample contain a clear
exponentially decaying part. We (as well as others before us)
propose that the rapid decay phase of the X-ray light curves of
Long GRBs is of internal origin, i.e. it represents the rapid
decay of the central engine activity. As in the case of Ultra Long
GRBs, we propose to associate this rapid decay with the
exponential spindown of the central black hole. As a result, we
suggest that the duration of a GRB depends closely on the magnetic
flux accumulated on the event horizon. This in turn can be
directly associated, through flux conservation, with the surface
magnetic field of the progenitor star. We have to state that  in some
short GRBs the rapid decay phase is exponential, this needs careful work to be understood.

A final note on timescales is in order here. In all the bursts
with signs of exponential decay, the time that this decay stops is
defined as $t_{\rm burst}$ (Zhang et al. 2014) which, in general, has nothing to do
with our decay timescale $\tau_{\rm BZ}$. The central engine may
continue to give off energy, but a different emission mechanism
(e.g. external shocks) will eventually overpower its emission.

\section*{Acknowledgements}

This work made use of data supplied by the UK Swift Science Data
Centre at the University of Leicester, and was supported by the
General Secretariat for Research and Technology of Greece and the
European Social Fund in the framework of Action `Excellence'.

{}

\section*{Appendix}

The electromagnetic black hole spindown is conceptually similar to
the spindown of the so-called Faraday disk, a conducting disk of
radius $r$, mass $M$ and angular velocity $\Omega$ threaded by a
certain amount of magnetic flux $\Psi_m$ and magnetic field $B$
(Fig.~6). The magnetic field is not generated by the disk itself,
but is generated and held in place by an external magnet. If we
assume the existence of a conducting path for electric currents to
close over the surface of the disk, the spindown rate is
proportional to $-\Psi_m^2 \Omega^2$ , and the disk loses
rotational kinetic energy at a rate proportional to $M r^2
\Omega\dot{\Omega}$. Equating the latter two expressions, we
deduce that the Faraday disk spins down exponentially as $
\Omega(t)\propto e^{-t/2\tau}$, and thus loses rotational energy
at a rate
\begin{equation}
\dot{E}(t)\propto e^{-t/\tau}\ .
\end{equation}
The characteristic decay timescale $\tau$ is proportional to the
mass and the square of the radius, and inversely proportional to
the square of the total accumulated magnetic flux, namely
\begin{equation}
\tau\propto M r^2/\Psi_m^2\propto M/r^2 B^2\ .
\end{equation}

\begin{figure}
\centering
\begin{minipage}[b]{1\linewidth}
\includegraphics[trim=0cm 0cm 0cm 0cm,
clip=true, width=10cm, angle=0]{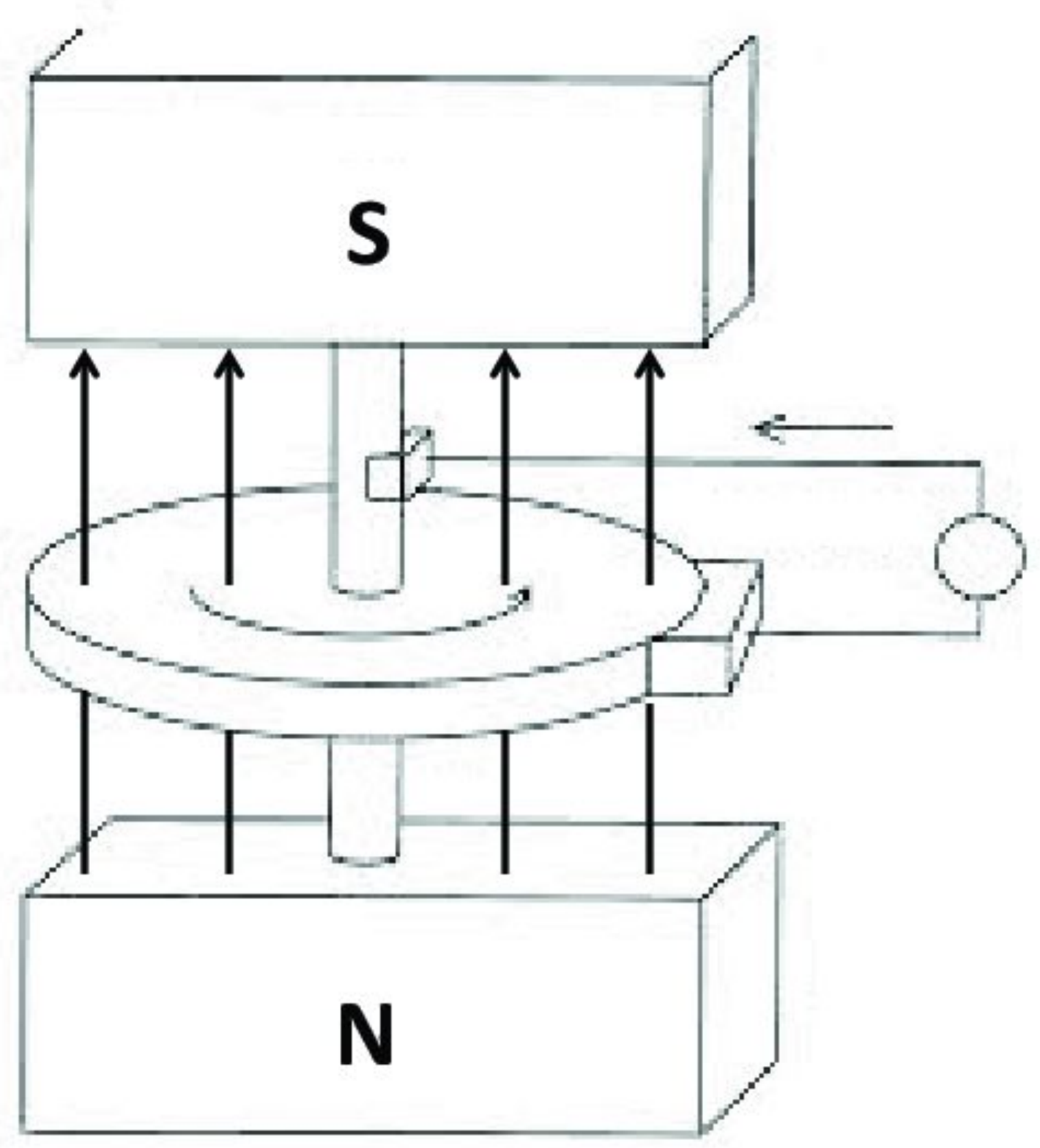}
\end{minipage}
\caption{Faraday disk with conducting path and load that allow it
to spin down exponentially. Vertical arrows: magnetic field.}
\end{figure}

\end{document}